\documentclass[epj,nopacs]{svjour}
\usepackage{cite}
\usepackage{epsf}
\usepackage{rotating}
\usepackage{here}

\def\kevc1{\ifmmode\mathrm{\ keV/{\mit c}}
          \else$\mathrm{\ keV/{\mit c}}$\fi}
\def\Mevc1{\ifmmode\mathrm{\ MeV/{\mit c}}
          \else$\mathrm{\ MeV/{\mit c}}$\fi}
\def\gevc1{\ifmmode\mathrm{ GeV/{\mit c}}
          \else$\mathrm{ GeV/{\mit c}}$\fi}
\def\kevc2{\ifmmode\mathrm{\ keV/{\mit c}^2}
          \else$\mathrm{\ keV/{\mit c}^2}$\fi}
\def\Mevc2{\ifmmode\mathrm{\ MeV/{\mit c}^2}
          \else$\mathrm{\ MeV/{\mit c}^2}$\fi}
\def\Gevc2{\ifmmode\mathrm{\ GeV/{\mit c}^2}
          \else$\mathrm{\ GeV/{\mit c}^2}$\fi}
\def\Gev2c2{\ifmmode\mathrm{\ GeV^2/{\mit c}^2}
          \else$\mathrm{\ GeV^2/{\mit c}^2}$\fi}

\newcommand{\dcy}{\mbox{$ \rightarrow $}}

\def\ubar{\ifmmode\mathrm{\overline {u}}
          \else$\mathrm{\overline{u}}$\fi}
\def\dbar{\ifmmode\mathrm{\overline {d}}
          \else$\mathrm{\overline{d}}$\fi}
\def\sbar{\ifmmode\mathrm{\overline {s}}
          \else$\mathrm{\overline{s}}$\fi}
\def\cbar{\ifmmode\mathrm{\overline {c}}
          \else$\mathrm{\overline{c}}$\fi}
\def\bbar{\ifmmode\mathrm{\overline {b}}
          \else$\mathrm{\overline{b}}$\fi}
\def\tbar{\ifmmode\mathrm{\overline {t}}
          \else$\mathrm{\overline{t}}$\fi}
\def\qbar{\ifmmode\mathrm{\overline {q}}
          \else$\mathrm{\overline{q}}$\fi}

\def\uq{\ifmmode\mathrm{u}
          \else$\mathrm{u}$\fi}
\def\dq{\ifmmode\mathrm{d}
          \else$\mathrm{d}$\fi}
\def\sq{\ifmmode\mathrm{s}
          \else$\mathrm{s}$\fi}
\def\cq{\ifmmode\mathrm{c}
          \else$\mathrm{c}$\fi}
\def\bq{\ifmmode\mathrm{b}
          \else$\mathrm{b}$\fi}
\def\tq{\ifmmode\mathrm{t}
          \else$\mathrm{t}$\fi}
\def\qq{\ifmmode\mathrm{q}
          \else$\mathrm{q}$\fi}

 \def\Pgg{\ifmmode\mathrm{\gamma}
          \else$\mathrm{\gamma}$\fi}
 \def\PW{\ifmmode\mathrm{W}
         \else$\mathrm{W }$\fi}
 \def\PWp{\ifmmode\mathrm{W^+}
          \else$\mathrm{W^+}$\fi}
 \def\PWpm{\ifmmode\mathrm{W^{\pm}}
          \else$\mathrm{W^{\pm}}$\fi}
 \def\PWm{\ifmmode\mathrm{W^-}
          \else$\mathrm{W^-}$\fi}
 \def\PZz{\ifmmode\mathrm{Z^0}
          \else$\mathrm{Z^0}$\fi}
 \def\PHz{\ifmmode\mathrm{H^0}
          \else$\mathrm{H^0}$\fi}
 \def\PHpm{\ifmmode\mathrm{H^{\pm}}
           \else$\mathrm{H^{\pm}}$\fi}
 \def\PWR{\ifmmode\mathrm{W_R}
          \else$\mathrm{W_R}$\fi}
 \def\PWpr{\ifmmode\mathrm{W^{\prime}}
           \else$\mathrm{W^{\prime}}$\fi}
 \def\PZLR{\ifmmode\mathrm{Z_{LR}}
           \else$\mathrm{Z_{LR}}$\fi}
 \def\PZgc{\ifmmode\mathrm{Z_{\chi}}
           \else$\mathrm{Z_{\chi}}$\fi}
 \def\PZgy{\ifmmode\mathrm{Z_{\psi}}
           \else$\mathrm{Z_{\psi}}$\fi}
 \def\PZge{\ifmmode\mathrm{Z_{\eta}}
           \else$\mathrm{Z_{\eta}}$\fi}
 \def\PZi{\ifmmode\mathrm{Z_1}
          \else$\mathrm{Z_1}$\fi}
 \def\PAz{\ifmmode\mathrm{A^0}
          \else$\mathrm{A^0}$\fi}
 \def\Pgne{\ifmmode\mathrm{\nu_{e}}
           \else$\mathrm{\nu_{e}}$\fi}
 \def\Pagne{\ifmmode\mathrm{\overline{\nu_{e}}}
            \else$\mathrm{\overline{\nu_{e}}}$\fi}
 \def\Pgngm{\ifmmode\mathrm{\nu_{\mu}}
            \else$\mathrm{\nu_{\mu}}$\fi}
 \def\Pagngm{\ifmmode\mathrm{\overline{\nu}_{\mu}}
             \else$\mathrm{\overline{\nu}_{\mu}}$\fi}
 \def\Pgngt{\ifmmode\mathrm{\nu_{\tau}}
            \else$\mathrm{\nu_{\tau}}$\fi}
 \def\Pagngt{\ifmmode\mathrm{\overline{\nu}_{\tau}}
             \else$\mathrm{\overline{\nu}_{\tau}}$\fi}
 \def\Pe{\ifmmode\mathrm{e}
         \else$\mathrm{e}$\fi}
 \def\Pep{\ifmmode\mathrm{e^+}
          \else$\mathrm{e^+}$\fi}
 \def\Pem{\ifmmode\mathrm{e^-}
          \else$\mathrm{e^-}$\fi}
 \def\Pgm{\ifmmode\mathrm{\mu}
          \else$\mathrm{\mu}$\fi}
 \def\Pgmm{\ifmmode\mathrm{\mu^-}
           \else$\mathrm{\mu^-}$\fi}
 \def\Pgmp{\ifmmode\mathrm{\mu^+}
           \else$\mathrm{\mu^+}$\fi}
 \def\Pgt{\ifmmode\mathrm{\tau}
          \else$\mathrm{\tau}$\fi}
 \def\PLpm{\ifmmode\mathrm{L^{\pm}}
           \else$\mathrm{L^{\pm}}$\fi}
 \def\PLz{\ifmmode\mathrm{L^0}
          \else$\mathrm{L^0}$\fi}
 \def\PEz{\ifmmode\mathrm{E^0}
          \else$\mathrm{E^0}$\fi}
 \def\Pgp{\ifmmode\mathrm{\pi}
          \else$\mathrm{\pi }$\fi}
 \def\Pgpm{\ifmmode\mathrm{\pi^-}
           \else$\mathrm{\pi^-}$\fi}
 \def\Pgpp{\ifmmode\mathrm{\pi^+}
           \else$\mathrm{\pi^+}$\fi}
 \def\Pgppm{\ifmmode\mathrm{\pi^{\pm }}
            \else$\mathrm{\pi^{\pm }}$\fi}
 \def\Pgpz{\ifmmode\mathrm{\pi^0}
           \else$\mathrm{\pi^0 }$\fi}
 \def\Pgh{\ifmmode\mathrm{\eta}
          \else$\mathrm{\eta }$\fi}
 \def\Pgr{\ifmmode\mathrm{\rho(770)}
          \else$\mathrm{\rho(770)}$\fi}
 \def\Pgo{\ifmmode\mathrm{\omega(783)}
          \else$\mathrm{\omega(783)}$\fi}
 \def\Pghpr{\ifmmode\mathrm{\eta^{\prime}(958)}
            \else$\mathrm{\eta^{\prime}(958)}$\fi}
 \def\Pfz{\ifmmode\mathrm{f_0(980)}
          \else$\mathrm{f_0(980)}$\fi}
 \def\Paz{\ifmmode\mathrm{a_0(980)}
          \else$\mathrm{a_0(980)}$\fi}
 \def\Pgf{\ifmmode\mathrm{\phi(1020)}
          \else$\mathrm{\phi(1020)}$\fi}
 \def\Phia{\ifmmode\mathrm{h_1(1170)}
           \else$\mathrm{h_1(1170)}$\fi}
 \def\Pbi{\ifmmode\mathrm{b_1(1235)}
          \else$\mathrm{b_1(1235)}$\fi}
 \def\Pai{\ifmmode\mathrm{a_1(1260)}
          \else$\mathrm{a_1(1260)}$\fi}
 \def\Pfii{\ifmmode\mathrm{f_2(1270)}
           \else$\mathrm{f_2(1270)}$\fi}
 \def\Pfi{\ifmmode\mathrm{f_1(1285)}
          \else$\mathrm{f_1(1285)}$\fi}
 \def\Pgha{\ifmmode\mathrm{\eta(1295)}
           \else$\mathrm{\eta(1295)}$\fi}
 \def\Pgpa{\ifmmode\mathrm{\pi(1300)}
           \else$\mathrm{\pi(1300)}$\fi}
 \def\Paii{\ifmmode\mathrm{a_2(1320)}
           \else$\mathrm{a_2(1320)}$\fi}
 \def\Pgoa{\ifmmode\mathrm{\omega(1390)}
           \else$\mathrm{\omega(1390)}$\fi}
 \def\Pfza{\ifmmode\mathrm{f_0(1400)}
           \else$\mathrm{f_0(1400)}$\fi}
 \def\Pfia{\ifmmode\mathrm{f_1 (1390)}
           \else$\mathrm{f_1 (1390)}$\fi}
 \def\Pghb{\ifmmode\mathrm{\eta(1440)}
           \else$\mathrm{\eta(1440)}$\fi}
 \def\Pgra{\ifmmode\mathrm{\rho(1450)}
           \else$\mathrm{\rho(1450)}$\fi}
 \def\Pfib{\ifmmode\mathrm{f_1(1510)}
           \else$\mathrm{f_1(1510)}$\fi}
 \def\Pfiipr{\ifmmode\mathrm{f^{\prime}_2(1525)}
             \else$\mathrm{f^{\prime}_2(1525)}$\fi}
 \def\Pfzb{\ifmmode\mathrm{f_0(1590)}
           \else$\mathrm{f_0(1590)}$\fi}
 \def\Pgob{\ifmmode\mathrm{\omega(1600)}
           \else$\mathrm{\omega(1600)}$\fi}
 \def\Pgoiii{\ifmmode\mathrm{\omega_3(1670)}
             \else$\mathrm{\omega_3(1670)}$\fi}
 \def\Pgpii{\ifmmode\mathrm{\pi_2(1670)}
            \else$\mathrm{\pi_2(1670)}$\fi}
 \def\Pgfa{\ifmmode\mathrm{\phi(1680)}
           \else$\mathrm{\phi(1680)}$\fi}
 \def\Pgriii{\ifmmode\mathrm{\rho_3(1690)}
             \else$\mathrm{\rho_3(1690)}$\fi}
 \def\Pgrb{\ifmmode\mathrm{\rho(1700)}
           \else$\mathrm{\rho(1700)}$\fi}
 \def\Pfiia{\ifmmode\mathrm{f_2(1720)}
            \else$\mathrm{f_2(1720)}$\fi}
 \def\Pgfiii{\ifmmode\mathrm{\phi_3(1850)}
             \else$\mathrm{\phi_3(1850)}$\fi}
 \def\Pfiib{\ifmmode\mathrm{f_2(2010)}
            \else$\mathrm{f_2(2010)}$\fi}
 \def\Pfiv{\ifmmode\mathrm{f_4(2050)}
           \else$\mathrm{f_4(2050)}$\fi}
 \def\Pfiic{\ifmmode\mathrm{f_2(2300)}
            \else$\mathrm{f_2(2300)}$\fi}
 \def\Pfiid{\ifmmode\mathrm{f_2(2340)}
            \else$\mathrm{f_2(2340)}$\fi}
 \def\PK{\ifmmode\mathrm{K}
         \else$\mathrm{K}$\fi}
 \def\PKpm{\ifmmode\mathrm{K^{\pm}}
           \else$\mathrm{K^{\pm}}$\fi}
 \def\PKp{\ifmmode\mathrm{K^+}
          \else$\mathrm{K^+}$\fi}
 \def\PKm{\ifmmode\mathrm{K^-}
          \else$\mathrm{K^-}$\fi}
 \def\PKz{\ifmmode\mathrm{K^0}
          \else$\mathrm{K^0}$\fi}
 \def\PaKz{\ifmmode\mathrm{\overline{K^0}}
           \else$\mathrm{\overline{K^0}}$\fi}
 \def\PKgmiii{\ifmmode\mathrm{K_{\mu 3}}
              \else$\mathrm{K_{\mu 3}}$\fi}
 \def\PKeiii{\ifmmode\mathrm{K_{\rm e3}}
             \else$\mathrm{K_{\rm e3}}$\fi}
 \def\PKzS{\ifmmode\mathrm{K^0_{\rm S}}
           \else$\mathrm{K^0_{\rm S}}$\fi}
 \def\PKzL{\ifmmode\mathrm{K^0_{\rm L}}
           \else$\mathrm{K^0_{\rm L}}$\fi}
 \def\PKzgmiii{\ifmmode\mathrm{K^0_{\mu 3}}
               \else$\mathrm{K^0_{\mu 3}}$\fi}
 \def\PKzeiii{\ifmmode\mathrm{K^0_{{\rm e}3}}
              \else$\mathrm{K^0_{{\rm e}3}}$\fi}
 \def\PKst{\ifmmode\mathrm{K^{\ast}(892)}
           \else$\mathrm{K^{\ast}(892)}$\fi}
 \def\PKi{\ifmmode\mathrm{K_1(1270)}
          \else$\mathrm{K_1(1270)}$\fi}
 \def\PKsta{\ifmmode\mathrm{K^{\ast}(1370)}
            \else$\mathrm{K^{\ast}(1370)}$\fi}
 \def\PKia{\ifmmode\mathrm{K_1(1400)}
           \else$\mathrm{K_1(1400)}$\fi}
 \def\PKstz{\ifmmode\mathrm{K^{\ast}_0(1430)}
            \else$\mathrm{K^{\ast}_0(1430)}$\fi}
 \def\PKstii{\ifmmode\mathrm{K^{\ast}_2(1430)}
             \else$\mathrm{K^{\ast}_2(1430)}$\fi}
 \def\PKstb{\ifmmode\mathrm{K^{\ast}(1680)}
            \else$\mathrm{K^{\ast}(1680)}$\fi}
 \def\PKii{\ifmmode\mathrm{K_2(1770)}
           \else$\mathrm{K_2(1770)}$\fi}
 \def\PKstiii{\ifmmode\mathrm{K^{\ast}_3(1780)}
              \else$\mathrm{K^{\ast}_3(1780)}$\fi}
 \def\PKstiv{\ifmmode\mathrm{K^{\ast}_4(2045)}
             \else$\mathrm{K^{\ast}_4(2045)}$\fi}
 \def\PD{\ifmmode\mathrm{D}
           \else$\mathrm{D}$\fi}
 \def\PaD{\ifmmode\mathrm{\overline{ D}}
          \else${\mathrm{\overline D}}$\fi}
 \def\PDpm{\ifmmode\mathrm{D^{\pm}}
           \else$\mathrm{D^{\pm}}$\fi}
 \def\PDm{\ifmmode\mathrm{D^-}
          \else$\mathrm{D^-}$\fi}
 \def\PDp{\ifmmode\mathrm{D^+}
          \else$\mathrm{D^+}$\fi}
 \def\PDz{\ifmmode\mathrm{D^0}
          \else$\mathrm{D^0}$\fi}
 \def\PaDz{\ifmmode\mathrm{\overline{D^0}}
           \else$\mathrm{\overline{D^0}}$\fi}
 \def\PDstpm{\ifmmode{\mathrm{D}^{\ast}(2010)^{\pm}}
             \else$\mathrm{D}^{\ast}(2010)^{\pm}$\fi}
 \def\PDstp{\ifmmode{\mathrm{D}^{\ast+}}
             \else$\mathrm{D}^{\ast+}$\fi}
 \def\PDst{\ifmmode{\mathrm{D}^{\ast}}
             \else$\mathrm{D}^{\ast}$\fi}
 \def\PDstz{\ifmmode{\mathrm{D}^{\ast}(2010)^0}
            \else$\mathrm{D}^{\ast}(2010)^0$\fi}
 \def\PDiz{\ifmmode{\mathrm{D}_{1}(2420)^0}
           \else$\mathrm{D}_{1}(2420)^0$\fi}
 \def\PDstiiz{\ifmmode{\mathrm{D}^{\ast}_{2}(2460)^0}
              \else$\mathrm{D}^{\ast}_{2}(2460)^0$\fi}
 \def\PsDp{\ifmmode\mathrm{D_{s}^+}
           \else$\mathrm{D_{s}^+}$\fi}
 \def\PsDm{\ifmmode\mathrm{D_{s}^-}
           \else$\mathrm{D_{s}^-}$\fi}
 \def\PsDpm{\ifmmode{\mathrm D_{s}^{\pm}}
           \else${\mathrm D_{s}^{\pm}}$\fi}
 \def\PsDst{\ifmmode\mathrm{D_{s}^{\ast}}
            \else$\mathrm{D_{s}^{\ast}}$\fi}
 \def\PsDipm{\ifmmode\mathrm{D_{s1}(2536)^{\pm}}
           \else$\mathrm{D_{s1}(2536)^{\pm}}$\fi}
 \def\PB{\ifmmode{\mathrm{B}}
          \else$\mathrm{B}$\fi}
 \def\PBp{\ifmmode{\mathrm{B}^{+}}
           \else$\mathrm{B}^{+}$\fi}
 \def\PBm{\ifmmode{\mathrm{B}^{-}}
           \else$\mathrm{B}^{-}$\fi}
 \def\PBpm{\ifmmode{\mathrm{B}^{\pm}}
            \else$\mathrm{B}^{\pm}$\fi}
 \def\PBz{\ifmmode{\mathrm{B}^0}
           \else$\mathrm{B}^0$\fi}
 \def\PbgL{\ifmmode{\mathrm{\Lambda}_b}
           \else$\mathrm{\Lambda}_b$\fi}
 \def\Pcgh{\ifmmode\mathrm{{\eta}_{c}(1S)}
           \else$\mathrm{{\eta}_{c}(1S)}$\fi}
 \def\PJgyy{\ifmmode\mathrm{J /\psi}
           \else$\mathrm{J /\psi}$\fi}
 \def\PJgy{\ifmmode\mathrm{J /\psi(1S)}
           \else$\mathrm{J /\psi(1S)}$\fi}
 \def\Pcgcz{\ifmmode\mathrm{{\chi}_{c0}(1P)}
            \else$\mathrm{{\chi}_{c0}(1P)}$\fi}
 \def\Pcgci{\ifmmode\mathrm{{\chi}_{c1}(1P)}
            \else$\mathrm{{\chi}_{c1}(1P)}$\fi}
 \def\Pcgcii{\ifmmode\mathrm{{\chi}_{c2}(1P)}
             \else$\mathrm{{\chi}_{c2}(1P)}$\fi}
 \def\Pgy{\ifmmode\mathrm{\psi(2S)}
          \else$\mathrm{\psi(2S)}$\fi}
 \def\Pgya{\ifmmode\mathrm{\psi(3770)}
           \else$\mathrm{\psi(3770)}$\fi}
 \def\Pgyb{\ifmmode\mathrm{\psi(4040)}
           \else$\mathrm{\psi(4040)}$\fi}
 \def\Pgyc{\ifmmode\mathrm{\psi(4160)}
           \else$\mathrm{\psi(4160)}$\fi}
 \def\Pgyd{\ifmmode\mathrm{\psi(4415)}
           \else$\mathrm{\psi(4415)}$\fi}
 \def\PgU{\ifmmode\mathrm{\Upsilon(1S)}
          \else$\mathrm{\Upsilon(1S)}$\fi}
 \def\Pbgcz{\ifmmode\mathrm{{\chi}_{b0}(1P)}
            \else$\mathrm{{\chi}_{b0}(1P)}$\fi}
 \def\Pbgci{\ifmmode\mathrm{{\chi}_{b1}(1P)}
            \else$\mathrm{{\chi}_{b1}(1P)}$\fi}
 \def\Pbgcii{\ifmmode\mathrm{{\chi}_{b2}(1P)}
             \else$\mathrm{{\chi}_{b2}(1P)}$\fi}
 \def\PgUa{\ifmmode\mathrm{\Upsilon(2S)}
           \else$\mathrm{\Upsilon(2S)}$\fi}
 \def\Pbgcza{\ifmmode\mathrm{{\chi}_{b0}(2P)}
             \else$\mathrm{{\chi}_{b0}(2P)}$\fi}
 \def\Pbgcia{\ifmmode\mathrm{{\chi}_{b1}(2P)}
             \else$\mathrm{{\chi}_{b1}(2P)}$\fi}
 \def\Pbgciia{\ifmmode\mathrm{{\chi}_{b2}(2P)}
              \else$\mathrm{{\chi}_{b2}(2P)}$\fi}
 \def\PgUb{\ifmmode\mathrm{\Upsilon(3S)}
           \else$\mathrm{\Upsilon(3S)}$\fi}
 \def\PgUc{\ifmmode\mathrm{\Upsilon(4S)}
           \else$\mathrm{\Upsilon(4S)}$\fi}
 \def\PgUd{\ifmmode\mathrm{\Upsilon(10860)}
           \else$\mathrm{\Upsilon(10860)}$\fi}
 \def\PgUe{\ifmmode\mathrm{\Upsilon(11020)}
           \else$\mathrm{\Upsilon(11020)}$\fi}
 \def\Pp{\ifmmode\mathrm{p}
         \else$\mathrm{p}$\fi}
 \def\Pap{\ifmmode\mathrm{\overline{p}}
         \else$\mathrm{\overline{p}}$\fi}
 \def\Pn{\ifmmode\mathrm{n}
         \else$\mathrm{n}$\fi}
 \def\PNa{\ifmmode\mathrm{N(1440)P_{11}}
          \else$\mathrm{N(1440)P_{11}}$\fi}
 \def\PNb{\ifmmode\mathrm{N(1520)D_{13}}
          \else$\mathrm{N(1520)D_{13}}$\fi}
 \def\PNc{\ifmmode\mathrm{N(1535)S_{11}}
          \else$\mathrm{N(1535)S_{11}}$\fi}
 \def\PNd{\ifmmode\mathrm{N(1650)S_{11}}
          \else$\mathrm{N(1650)S_{11}}$\fi}
 \def\PNe{\ifmmode\mathrm{N(1675)D_{15}}
          \else$\mathrm{N(1675)D_{15}}$\fi}
 \def\PNf{\ifmmode\mathrm{N(1680)F_{15}}
          \else$\mathrm{N(1680)F_{15}}$\fi}
 \def\PNg{\ifmmode\mathrm{N(1700)D_{13}}
          \else$\mathrm{N(1700)D_{13}}$\fi}
 \def\PNh{\ifmmode\mathrm{N(1710)P_{11}}
          \else$\mathrm{N(1710)P_{11}}$\fi}
 \def\PNi{\ifmmode\mathrm{N(1720)P_{13}}
          \else$\mathrm{N(1720)P_{13}}$\fi}
 \def\PNj{\ifmmode\mathrm{N(2190)G_{17}}
          \else$\mathrm{N(2190)G_{17}}$\fi}
 \def\PNk{\ifmmode\mathrm{N(2220)H_{19}}
          \else$\mathrm{N(2220)H_{19}}$\fi}
 \def\PNl{\ifmmode\mathrm{N(2250)G_{19}}
          \else$\mathrm{N(2250)G_{19}}$\fi}
 \def\PNm{\ifmmode\mathrm{N(2600)I_{1,11}}
          \else$\mathrm{N(2600)I_{1,11}}$\fi}
 \def\PgDa{\ifmmode\mathrm{\Delta(1232)P_{33}}
           \else$\mathrm{\Delta(1232)P_{33}}$\fi}
 \def\PgDb{\ifmmode\mathrm{\Delta(1620)S_{31}}
           \else$\mathrm{\Delta(1620)S_{31}}$\fi}
 \def\PgDc{\ifmmode\mathrm{\Delta(1700)D_{33}}
           \else$\mathrm{\Delta(1700)D_{33}}$\fi}
 \def\PgDd{\ifmmode\mathrm{\Delta(1900)S_{31}}
           \else$\mathrm{\Delta(1900)S_{31}}$\fi}
 \def\PgDe{\ifmmode\mathrm{\Delta(1905)F_{35}}
           \else$\mathrm{\Delta(1905)F_{35}}$\fi}
 \def\PgDf{\ifmmode\mathrm{\Delta(1910)P_{31}}
           \else$\mathrm{\Delta(1910)P_{31}}$\fi}
 \def\PgDh{\ifmmode\mathrm{\Delta(1920)P_{33}}
           \else$\mathrm{\Delta(1920)P_{33}}$\fi}
 \def\PgDi{\ifmmode\mathrm{\Delta(1930)D_{35}}
           \else$\mathrm{\Delta(1930)D_{35}}$\fi}
 \def\PgDj{\ifmmode\mathrm{\Delta(1950)F_{37}}
           \else$\mathrm{\Delta(1950)F_{37}}$\fi}
 \def\PgDk{\ifmmode\mathrm{\Delta(2420)H_{3,11}}
           \else$\mathrm{\Delta(2420)H_{3,11}}$\fi}
 \def\PgDpp{\ifmmode\mathrm{\Delta^{++}}
           \else$\mathrm{\Delta^{++}}$\fi}
 \def\PgL{\ifmmode\mathrm{\Lambda}
          \else$\mathrm{\Lambda}$\fi}
 \def\PgLa{\ifmmode\mathrm{\Lambda(1405) S_{01}}
           \else$\mathrm{\Lambda(1405) S_{01}}$\fi}
 \def\PgLb{\ifmmode\mathrm{\Lambda(1520) D_{03}}
           \else$\mathrm{\Lambda(1520) D_{03}}$\fi}
 \def\PgLc{\ifmmode\mathrm{\Lambda(1600) P_{01}}
           \else$\mathrm{\Lambda(1600) P_{01}}$\fi}
 \def\PgLd{\ifmmode\mathrm{\Lambda(1670) S_{01}}
           \else$\mathrm{\Lambda(1670) S_{01}}$\fi}
 \def\PgLe{\ifmmode\mathrm{\Lambda(1690) D_{03}}
           \else$\mathrm{\Lambda(1690) D_{03}}$\fi}
 \def\PgLf{\ifmmode\mathrm{\Lambda(1800) S_{01}}
           \else$\mathrm{\Lambda(1800) S_{01}}$\fi}
 \def\PgLg{\ifmmode\mathrm{\Lambda(1810) P_{01}}
           \else$\mathrm{\Lambda(1810) P_{01}}$\fi}
 \def\PgLh{\ifmmode\mathrm{\Lambda(1820) F_{05}}
           \else$\mathrm{\Lambda(1820) F_{05}}$\fi}
 \def\PgLi{\ifmmode\mathrm{\Lambda(1830) D_{05}}
           \else$\mathrm{\Lambda(1830) D_{05}}$\fi}
 \def\PgLj{\ifmmode\mathrm{\Lambda(1890) P_{03}}
           \else$\mathrm{\Lambda(1890) P_{03}}$\fi}
 \def\PgLk{\ifmmode\mathrm{\Lambda(2100) G_{07}}
           \else$\mathrm{\Lambda(2100) G_{07}}$\fi}
 \def\PgLl{\ifmmode\mathrm{\Lambda(2110) F_{05}}
           \else$\mathrm{\Lambda(2110) F_{05}}$\fi}
 \def\PgLm{\ifmmode\mathrm{\Lambda(2350) H_{09}}
           \else$\mathrm{\Lambda(2350) H_{09}}$\fi}
 \def\PgS{\ifmmode{\rm \Sigma}
           \else${\rm \Sigma}$\fi}
 \def\PgSp{\ifmmode\mathrm{\Sigma^+}
           \else$\mathrm{\Sigma^+}$\fi}
 \def\PgSz{\ifmmode\mathrm{\Sigma^0}
           \else$\mathrm{\Sigma^0}$\fi}
 \def\PgSm{\ifmmode\mathrm{\Sigma^-}
           \else$\mathrm{\Sigma^-}$\fi}
 \def\PgSpm{\ifmmode\mathrm{\Sigma^{\pm}}
           \else$\mathrm{\Sigma^{\pm}}$\fi}
 \def\PgSa{\ifmmode\mathrm{\Sigma(1385) P_{13}}
           \else$\mathrm{\Sigma(1385) P_{13}}$\fi}
 \def\PgSb{\ifmmode\mathrm{\Sigma(1660) P_{11}}
           \else$\mathrm{\Sigma(1660) P_{11}}$\fi}
 \def\PgSc{\ifmmode\mathrm{\Sigma(1670) D_{13}}
           \else$\mathrm{\Sigma(1670) D_{13}}$\fi}
 \def\PgSd{\ifmmode\mathrm{\Sigma(1750) S_{11}}
           \else$\mathrm{\Sigma(1750) S_{11}}$\fi}
 \def\PgSe{\ifmmode\mathrm{\Sigma(1775) D_{15}}
           \else$\mathrm{\Sigma(1775) D_{15}}$\fi}
 \def\PgSf{\ifmmode\mathrm{\Sigma(1915) F_{15}}
           \else$\mathrm{\Sigma(1915) F_{15}}$\fi}
 \def\PgSg{\ifmmode\mathrm{\Sigma(1940) D_{13}}
           \else$\mathrm{\Sigma(1940) D_{13}}$\fi}
 \def\PgSh{\ifmmode\mathrm{\Sigma(2030) F_{17}}
           \else$\mathrm{\Sigma(2030) F_{17}}$\fi}
 \def\PgSi{\ifmmode\mathrm{\Sigma(2050)}
           \else$\mathrm{\Sigma(2050)}$\fi}
 \def\PgXz{\ifmmode\mathrm{\Xi^0}
           \else$\mathrm{\Xi^0}$\fi}
 \def\PgXm{\ifmmode\mathrm{\Xi^-}
           \else$\mathrm{\Xi^-}$\fi}
 \def\PgXa{\ifmmode\mathrm{\Xi(1530)}
           \else$\mathrm{\Xi(1530)}$\fi}
 \def\PgXas{\ifmmode\mathrm{\Xi(1530)P_{13}}
           \else$\mathrm{\Xi(1530)P_{13}}$\fi}
 \def\PgXb{\ifmmode\mathrm{\Xi(1690)}
           \else$\mathrm{\Xi(1690)}$\fi}
 \def\PgXbb{\ifmmode\mathrm{\Xi(1620)}
           \else$\mathrm{\Xi(1620)}$\fi}
 \def\PgXc{\ifmmode\mathrm{\Xi(1820)D_{13}}
           \else$\mathrm{\Xi(1820)D_{13}}$\fi}
 \def\PgXcs{\ifmmode\mathrm{\Xi(1820)}
           \else$\mathrm{\Xi(1820)}$\fi}
 \def\PgXd{\ifmmode\mathrm{\Xi(1950)}
           \else$\mathrm{\Xi(1950)}$\fi}
 \def\PgXe{\ifmmode\mathrm{\Xi(2030)}
           \else$\mathrm{\Xi(2030)}$\fi}
 \def\PgOm{\ifmmode\mathrm{\Omega^-}
           \else$\mathrm{\Omega^-}$\fi}
 \def\PgO{\ifmmode\mathrm{\Omega}
           \else$\mathrm{\Omega}$\fi}
 \def\PgOma{\ifmmode\mathrm{\Omega(2250)^-}
            \else$\mathrm{\Omega(2250)^-}$\fi}
 \def\PcgL{\ifmmode\mathrm{\Lambda_c}
            \else$\mathrm{\Lambda_c}$\fi}
 \def\PacgL{\ifmmode\mathrm{\overline{\Lambda}_c}
            \else$\mathrm{\overline{\Lambda}_c}$\fi}
 \def\PcgLp{\ifmmode\mathrm{\Lambda_c^+}
            \else$\mathrm{\Lambda_c^+}$\fi}
 \def\PcgLm{\ifmmode{\rm \Lambda_c^-}
            \else${\rm \Lambda_c^-}$\fi}
 \def\PcgX{\ifmmode\mathrm{\Xi_c}
            \else$\mathrm{\Xi_c}$\fi}
 \def\PcgXz{\ifmmode\mathrm{\Xi_c^0}
            \else$\mathrm{\Xi_c^0}$\fi}
 \def\PcgXp{\ifmmode\mathrm{\Xi_c^+}
            \else$\mathrm{\Xi_c^+}$\fi}
 \def\PcgS{\ifmmode\mathrm{\Sigma_c}
           \else$\mathrm{\Sigma_c}$\fi}
 \def\PcgSz{\ifmmode\mathrm{\Sigma_c^0}
           \else$\mathrm{\Sigma_c^0}$\fi}
 \def\PcgSp{\ifmmode\mathrm{\Sigma_c^+}
           \else$\mathrm{\Sigma_c^+}$\fi}
 \def\PcgSpp{\ifmmode\mathrm{\Sigma_c^{++}}
           \else$\mathrm{\Sigma_c^{++}}$\fi}
 \def\PcgO{\ifmmode{\mathrm \Omega_c}
           \else${\mathrm \Omega_c}$\fi}
 \def\PcgOz{\ifmmode{\mathrm \Omega_c^{0}}
           \else${\mathrm \Omega_c^{0}}$\fi}
 \def\PSgg{\ifmmode\mathrm{\tilde{\gamma}}
           \else$\mathrm{\tilde{\gamma}}$\fi}
 \def\PSgxz{\ifmmode\mathrm{\tilde{\chi}^0_i}
            \else$\mathrm{\tilde{\chi}^0_i}$\fi}
 \def\PSZz{\ifmmode\mathrm{\tilde{Z}^0}
           \else$\mathrm{\tilde{Z}^0}$\fi}
 \def\PSHz{\ifmmode\mathrm{\tilde{H}^0_j}
           \else$\mathrm{\tilde{H}^0_j}$\fi}
 \def\PSgxpm{\ifmmode\mathrm{\tilde{\chi}^{\pm_i}}
             \else$\mathrm{\tilde{\chi}^{\pm_i}}$\fi}
 \def\PSWpm{\ifmmode\mathrm{\tilde{W}^{\pm}}
            \else$\mathrm{\tilde{W}^{\pm}}$\fi}
 \def\PSHpm{\ifmmode\mathrm{\tilde{H}^{\pm_j}}
            \else$\mathrm{\tilde{H}^{\pm_j}}$\fi}
 \def\PSgn{\ifmmode\mathrm{\tilde{\nu}}
           \else$\mathrm{\tilde{\nu}}$\fi}
 \def\PSe{\ifmmode\mathrm{\tilde{e}}
          \else$\mathrm{\tilde{e}}$\fi}
 \def\PSgm{\ifmmode\mathrm{\tilde{\mu}}
           \else$\mathrm{\tilde{\mu}}$\fi}
 \def\PSgt{\ifmmode\mathrm{\tilde{\tau}}
           \else$\mathrm{\tilde{\tau}}$\fi}
 \def\PSq{\ifmmode\mathrm{\tilde{q}}
          \else$\mathrm{\tilde{q}}$\fi}
 \def\PSg{\ifmmode\mathrm{\tilde{g}}
          \else$\mathrm{\tilde{g}}$\fi}

\begin{document}

\hugehead 
\newcounter{str1}
\newcounter{str2}
\newcounter{str3}
\newcounter{str4}
\newcounter{str5}

\title{
Determination of the Total $c\bar{c}$ Production Cross Section in 340 \gevc1\ 
$\PgSm$-Nucleus Interactions} 
\titlerunning{~}
\authorrunning{~}
\subtitle{The WA89 Collaboration}
\author{
\renewcommand{\thefootnote}{{\rm\alph{footnote}}}%
M.I.~Adamovich         \inst{8}  \and 
Yu.A.~Alexandrov       \inst{8}  \fnmsep
 \thanks{supported by the Deutsche Forschungsgemeinschaft,
           contract number436 RUS 113/465, and Russian Foundation for
           Basic Research under contract number RFFI 98-02-04096.}
                     \setcounter{str5}{\value{footnote}} 
 \and
D.~Barberis            \inst{3}  \and 
M.~Beck                \inst{5}  \and 
C.~B\'erat             \inst{4}  \and 
W.~Beusch              \inst{2}  \and 
M.~Boss                \inst{6}  \and 
S.~Brons               \inst{5} \fnmsep \thanks{Now at TRIUMF, Vancouver, B.C., Canada V6T 2A3}  \and 
W.~Br\"uckner          \inst{5}  \and 
M.~Bu\'enerd           \inst{4}  \and 
C.~Busch               \inst{6}  \and 
C.~B\"uscher           \inst{5}  \and 
F.~Charignon           \inst{4}  \and 
J.~Chauvin             \inst{4}  \and 
E.A.~Chudakov          \inst{6} \fnmsep 
\thanks{Now at Thomas Jefferson Lab, Newport News, VA 23606, USA} 
                                \setcounter{str3}{\value{footnote}}\and
U.~Dersch              \inst{5}  \and F.~Dropmann            \inst{5}  \and J.~Engelfried          
\inst{6} \fnmsep \thanks{Now at Institudo de Fisica, Universidad San Luis Potosi, S.L.P. 
78240, Mexico} \and F.~Faller              \inst{6} \fnmsep \thanks{Now at Fraunhofer 
Institut f\"ur Solarenergiesysteme, D-79100 Freiburg, Germany} \and A.~Fournier            
\inst{4} \and S.G.~Gerassimov        \inst{5} \fnmsep \inst{8} \fnmsep \thanks{Now at 
Technische Universit\"at M\"unchen, Garching, Germany} 
                                \setcounter{str2}{\value{footnote}} \and
M.~Godbersen           \inst{5} \and 
P.~Grafstr\"om         \inst{2} \and 
Th.~Haller             \inst{5} \and 
M.~Heidrich            \inst{5} \and 
E.~Hubbard             \inst{5} \and 
R.B.~Hurst             \inst{3} \and 
K.~K\"onigsmann        \inst{5} \fnmsep 
\thanks{Now at Fakult\"at f\"ur Physik, Universit\"at Freiburg, Germany} 
                                \setcounter{str1}{\value{footnote}} \and
I.~Konorov             \inst{5} \fnmsep \inst{8} \fnmsep \footnotemark[\value{str2}] \and 
N.~Keller              \inst{6}  \and K.~Martens             \inst{6} \fnmsep \thanks{ 
Now at Department of Physics and Astronomy, SUNY at Stony Brook, NY 11794-3800, USA} \and 
Ph.~Martin             \inst{4} \and S.~Masciocchi          \inst{5} \fnmsep 
\thanks{Now at Max-Planck-Institut f\"ur Physik, M\"unchen, Germany} \and 
R.~Michaels            \inst{5} \fnmsep \footnotemark[\value{str3}]  \and U.~M\"uller            
\inst{7}  \and H.~Neeb                \inst{5}  \and D.~Newbold             \inst{1}  
\and C.~Newsom              \thanks{University of Iowa, Iowa City, IA 52242, USA} \and 
S.~Paul                \inst{5} \fnmsep \footnotemark[\value{str2}] \and J.~Pochodzalla         
\inst{5}  \and I.~Potashnikova        \inst{5}  \and B.~Povh                \inst{5}  
\and R.~Ransome             \thanks{Rutgers University, Piscataway, NJ 08854, USA.} \and 
Z.~Ren                 \inst{5}  \and M.~Rey-Campagnolle     \inst{4} \fnmsep 
\thanks{permanent address: CERN, CH-1211 Gen\`eve 23, Switzerland}  \and G.~Rosner              
\inst{7}  \and L.~Rossi               \inst{3}  \and H.~Rudolph             \inst{7}  
\and C.~Scheel              \thanks{NIKHEF, 1009 DB Amsterdam, The Netherlands}  \and 
L.~Schmitt             \inst{7} \fnmsep \footnotemark[\value{str2}] \and H.-W.~Siebert          
\inst{6}  \and A.~Simon               \inst{6} \fnmsep \footnotemark[\value{str1}] \and 
V.J.~Smith             \inst{1} \fnmsep \thanks{supported by the UK PPARC}  \and 
O.~Thilmann            \inst{6}  \and A.~Trombini            \inst{5}  \and E.~Vesin               
\inst{4}  \and B.~Volkemer            \inst{7}  \and K.~Vorwalter           \inst{5}  
\and Th.~Walcher            \inst{7}  \and G.~W\"alder            \inst{6}  \and 
R.~Werding             \inst{5}  \and E.~Wittmann            \inst{5}  \and 
M.V.~Zavertyaev        \inst{8}  \fnmsep \footnotemark[\value{str5}]  } 

\institute{ 
\renewcommand{\thefootnote}{{\rm\alph{footnote}}}%
             University of Bristol, Bristol, United Kingdom \and
             CERN, CH-1211 Gen\`eve 23, Switzerland. \and
             Genoa University/INFN, Dipt. di Fisica,I-16146 Genova, Italy. \and
             Grenoble ISN, F-38026 Grenoble, France.  \and
             Max-Planck-Institut f\"ur Kernphysik, Postfach
             103980, D-69029 Heidelberg, Germany. \and
             Universit\"at Heidelberg, Physikal. Inst., D-69120 Heidelberg, Germany.
             \thanks{supported by the Bundesministerium f\"ur Bildung,
                     Wissenschaft, Forschung und Technologie,
                     Germany, under contract numbers 05~5HD15I, 06~HD524I and 06~MZ5265}
                     \setcounter{str4}{\value{footnote}} \and
             Universit\"at Mainz, Inst. f\"ur Kernphysik, D-55099 Mainz,
             Germany. \footnotemark[\value{str4}] \and
             Moscow Lebedev Physics Inst., RU-117924, Moscow, Russia.
\vspace*{.7cm}
              }
\abstract{ The production of charmed particles by \PgSm\ of $340\,\gevc1$ momentum was 
studied in the hyperon beam experiment WA89 at the CERN-SPS, using the 
$\Omega$-spectrometer. In two data-taking periods in 1993 and 1994 an integrated 
luminosity of $1600\,\mu\mbox{b}^{-1}$ on copper and carbon targets was recorded.  From 
the reconstruction of $ 930 \pm 90$ charm particle decays in 10 decay channels production 
cross sections for \PD , \PaD , \PsDm and \PcgLp\ were determined in the region $x_F>0$. 
Assuming an $A^1$ dependence of the cross section on the nucleon number, we calculate a 
total $c\bar{c}$ production cross section of $\sigma_{\cq\cbar}(x_F > 0) = 5.3\, \pm\, 
0.4\,\, (stat)\,\, \pm\, 1.0\,\, (syst)\,\, +1.0\,\,(\PcgX )\, \mu\mbox{b}$ per nucleon.  
The last term is an upper limit on the unknown contribution from charmed-strange baryon 
production.} 

\maketitle

\section{Introduction}

Nearly 25 years after the discovery of the charm quark, charm hadron
physics is still a major field of research. This results from the
difficult identification of charm particle decays due to low
production cross sections, short lifetimes and branching ratios of
only a few percent in the principal decay modes. Not only the
properties of the charmed hadrons themselves are the object of intense
experimental studies. Charm quark production is also of interest, since it
can be used for tests of QCD calculations. 
As the lightest of the ``heavy'' quarks,
the charm quark is of special interest as its production can be
described using perturbative methods and it is experimentally accessible
with reasonable statistics.

Usually charm particle production is modeled as a two-step process
involving different energy scales:
one hard part, which describes the production of the charm
quark pair itself, and the soft process of subsequent hadronization.

The elementary charm quark production process, dominated
at typical fixed target
energies by gluon-gluon fusion, has been calculated in the
framework of perturbative QCD in leading and next to leading order,
the NLO contributions exceeding the leading ones by a factor of
three (see, for instance, \cite{nason89}).

Because of the relatively small charm quark mass,  NLO calculations
still contain considerable uncertainties 
coming from uncertainties in the renormalization and factorization scales,
unprecise parton density functions as well as the unknown charm quark
mass.  
At present these uncertainties are 
much larger than the experimental errors.
Therefore measurements of charm production 
in different beams and at different energies provide
a data set on which many further developments of QCD-based
calculations can be tested rigorously.

In fixed target experiments charm particle production has so far been
studied mainly with pion and proton beams, with beam momenta ranging
from 200 to 600 and 200 to 800 \gevc1 , respectively. 
Only few data on charm
production by a kaon beam exist so far.
All measured cross sections lie within the
wide margins of the NLO calculations.
Many of these experiments measured
\PD /\PaD\ production cross sections only,
and the total charm production cross sections have to
be extrapolated from these measurements.

In this paper we report on the first measurement of the total $c\bar{c}$ cross section in 
\PgSm\ - nucleus interactions at $\sqrt{s} = 25\,\mbox{GeV/c}^2$, based on observed 
samples of \PDpm , \PDz , \PaDz , \PsDm , and \PcgLp . Results on the charge asymmetries 
in the production of \PDp , \PDz , \PsDp , \PcgL\ and their antiparticles have already 
been published \cite{asym}. 

\begin{figure*}
\leavevmode
   \mbox{
     \epsfxsize=17cm
     \epsffile{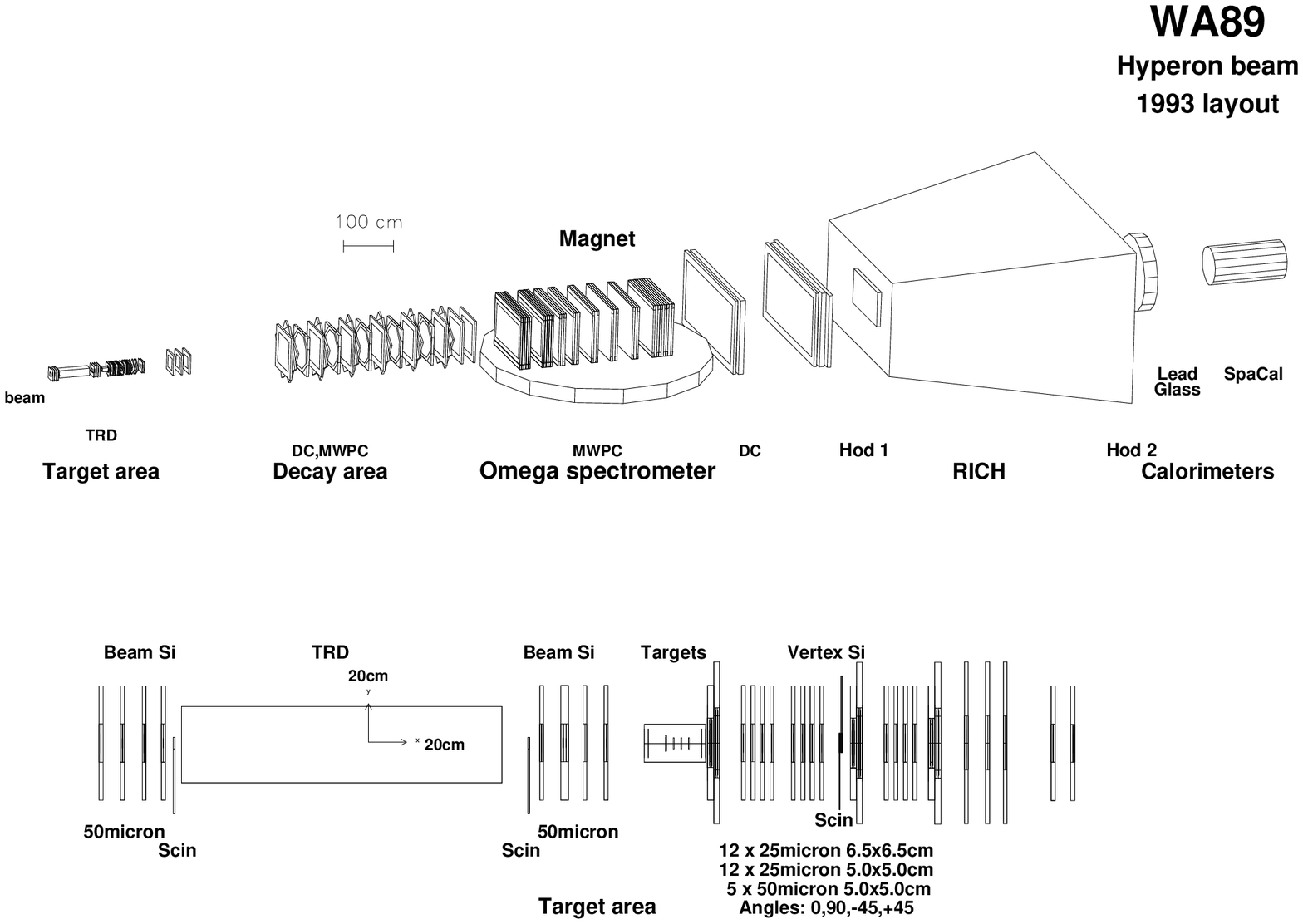}
     }
\caption{Setup of experiment WA89 in the 1993 run.
        The lower part shows an expanded view of the target area.}
\label{fig:wa89setup}
\end{figure*}

\section{Experimental setup}

Experiment WA89 at the CERN-SPS used a secondary beam of \PgSm\ created in 
$450\,\gevc1\,\,$ p-Be interactions. The detector was designed as a typical forward 
spectrometer, set up in the CERN West area around the $\Omega$-magnet, and consisted of a 
target region, a decay region, the $\Omega$-spectrometer, a ring imaging Cherenkov 
counter and an electromagnetic and a hadronic calorimeter. Figure \ref{fig:wa89setup} 
shows the setup in 1993 -- the setup in 1994 differed only margi\-nal\-ly. 

The hyperon beam was produced from $450\,\gevc1$ protons from the CERN-SPS impinging on a 
beryllium rod of $40\,\mbox{cm}$ length and $2\,\mbox{mm}$ diameter positioned 
$16\,\mbox{m}$ upstream of the experimental target. A magnetic channel selected negative 
particles with a mean momentum of 345 GeV/c and  a momentum spread of $\sigma(p)/p = 
9\%$.  A beam of $4.0\times 10^{10}$ protons per 2.1 seconds spill in a machine cycle of 
14.5 seconds yielded about $4.5\times 10^5\,\pi^-$ and $1.8\times 10^5\,\PgSm$ at the 
experimental target. A transition radiation detector was used to suppress \Pgpm\ on the 
trigger level \cite{trad}. A detailed description of the hyperon beam setup and 
parameters can be found in \cite{beam}. 

The experimental target consisted of one $4\,\mbox{mm}$
thick copper plate
and three $2.2\,\mbox{mm}$ thick carbon plates spaced at
$2\,\mbox{cm}$ intervals along the beam. For the carbon targets
industrial diamond with a density of $3.3\,\mbox{g/cm}^3$
was used instead of graphite since it allows for
thinner targets at the same interaction length, thus reducing the
decay probability inside the target. The total interaction length of
all targets was 4.95\%. The spacing of $2\,\mbox{cm}$
ensured that most charm decays occurred in the  gaps
between the targets and
hence had a lower background from secondary interactions.

A total of 23 silicon microstrip detectors with a pitch of
25 and $50\,\mu\mbox{m}$, assembled closely downstream of the targets,
was used for the reconstruction  of 
the charm production and decay vertices.
Another set of 14 planes upstream of the targets was used  for beam track
reconstruction~\cite{micro}.

The vertex area was followed by a $12\,\mbox{m}$ long decay zone for short-living strange 
particles, equipped with 38 drift chamber planes with a wire spacing of $5\,\mbox{cm}$, a 
sensitive area of $80\times 80\,\mbox{cm}^2$ and a spatial resolution of 
$300\,\mu\mbox{m}$. The mean efficiency was 88\%. To increase the tracking efficiency in 
the central region with its high track density the drift chambers were interleaved with 5 
sets of MWPCs of 4 planes each. These MWPCs had a wire spacing of $1\,\mbox{mm}$, a 
sensitive region of $\approx 12\times12\,\mbox{cm}^2$ and an efficiency of 99\%. 

To improve the track connection between the target and decay regions
three sets of 4 MWPCs with a wire spacing of $1\,\mbox{mm}$ were
installed \ $2\,\mbox{m}$ downstream of the target.

Particle momenta were measured in the $\Omega$-spectrometer, a
super-conducting magnet with an integrated field of $7.5\,\mbox{Tm}$.
A total of 45 MPWCs inside the magnet and 8 driftchamber planes and
4 MWPCs at the magnet exit allowed a momentum
resolution of $\sigma(p)/p^2 \approx10^{-4} (\gevc1)^{-1}$.

Charged particle identification was accomplished by a ring imaging
Cherenkov counter with a $5\,\mbox{m}$ long radiator volume filled with
nitrogen at atmospheric pressure. The Cherenkov threshold was at 
$\gamma=41$.  
UV photons were detected in driftchambers filled with TMAE-saturated
ethylene.
A resolution of $\sigma = 2.8\,\mbox{mm}$ and a mean
number of $15.5$ photoelectrons per ring, equivalent to a quality
factor of $N_0 = 53\,\mbox{cm}^{-1}$, allowed pion/kaon and pion/proton
separation up to a momentum of $90\,\gevc1$ and $150\,\gevc1$, resp.\
with a rejection factor of 10 \cite{rich}.

An electro-magnetic lead glass calorimeter and a hadro\-nic lead/scintillator 
``spaghetti'' calorimeter were located downstream of the RICH. Both were not used in the 
analysis presented here. 

The trigger was relatively open and had an average efficiency of
60-70\% for the charm decays discussed here. It required a minimum track
multiplicity in the target region, derived from two scintillators
placed after the first 12 microstrip planes. Also two high-momentum
particles were required with their momenta estimated from hit
correlations in scintillators and wire chambers inside and behind the
magnet.

The experiment had two main data taking periods in 1993 and 1994 during
which a total of 350 million interaction triggers were recorded,
corresponding to an integrated luminosity of $450\,\mu\mbox{b}^{-1}$
on copper and $1160\,\mu\mbox{b}^{-1}$ on carbon.

\section{Data Analysis}

Since there were only minor differences in the detector setups of  1993
and 1994, both data sets could be handled identically.
The reconstruction of charm particle decays followed a candidate-driven
approach: for each track combination
in the target area whose charges matched the decay considered,
reconstruction of the decay vertex was attempted.
If the reconstruction was successful,   
the production vertex was reconstructed from the remaining tracks
rejecting those which  contributed most 
to the $\chi^2$ of the vertex fit until
$\chi^2/(\mbox{degrees of freedom})$ fell below a given limit.

\begin{table*}[t]
  \begin{center}
  \begin{tabular}[h]{|lcl|c|c|c|c|}\hline
  \multicolumn{3}{|c|}{ Decay mode } &  branching ratio & events/background &
          n    & b [$c^2/GeV^2$]\\ \hline
 \PDz   & \dcy &\PKm\Pgpp           & $ (3.83\pm 0.12)\%$ & $ 140/54 $
                                    & $6.2\pm0.8$ & \\
        & \dcy &\PKm\Pgpp\Pgpp\Pgpm & $ (7.5 \pm 0.4) \%$ & $  86/29 $
                                    & $5.5\pm0.8$ & \\
 \PaDz  & \dcy &\PKp\Pgpm           & $ (3.83\pm 0.12)\%$ & $ 195/40 $
                                    & $4.5\pm0.4$ & \\
        & \dcy &\PKp\Pgpm\Pgpp\Pgpm & $ (7.5 \pm 0.4) \%$ & $  99/21 $
                                    & $4.6\pm0.6$ & \\
 \PDp   & \dcy &\PKm\Pgpp\Pgpp      & $ (9.1 \pm 0.6) \%$ & $ 116/30 $
                                    & $5.4\pm0.6$  & $1.15\pm0.15$ \\
 \PDm   & \dcy &\PKp\Pgpm\Pgpm      & $ (9.1 \pm 0.6) \%$ & $ 187/29 $
                                    & $4.2\pm0.4$ & $1.15\pm0.15$ \\
 \PsDm  & \dcy &\PKp\PKm\Pgpm       & $ (4.6 \pm 1.2) \%$ & $  55/15 $
                                    & $4.0\pm0.8$ & $0.79\pm0.12$ \\
 \PcgL  & \dcy &\PKm\Pgpp\Pp        & $ (4.4 \pm 0.6) \%$ & $  49/15 $
                                    & $3.0\pm1.0$ & $0.78\pm0.20$ \\ \hline
  \end{tabular}
  \caption{Numbers of observed events and values of  the production
    parameters $n$ and $b$ in the decay channels analyzed.
    Branching ratios are taken from \cite{pdg96}.}
  \label{tab:channels}
\end{center}
\end{table*}

Opposed to the total analysed sample which contains about 350 million events,
the final charm samples show a total of only a few 100 entries per
histogram. The reduction factor of more than $10^6$ was accomplished
in three passes by cuts
on (values in brackets give typical cut values and resolutions):
\begin{itemize}
  \item the separation between production and decay vertex
        $(>7-13\,\sigma ;\,\sigma\approx 500\,\mu\mbox{m})$,
  \item the impact parameter of the reconstructed charm candidate track 
        w.r.t. the production vertex $(<3-6\,\sigma ;\,\sigma\approx
        10\,\mu\mbox{m})$,
  \item the impact parameters of the decay vertex tracks w.r.t. the
        production vertex $(>3-6\,\sigma ;\,\sigma\approx
        10\,\mu\mbox{m})$
\end{itemize}
Additional requirements were:
\begin{itemize}
  \item a decay vertex location outside the targets
  \item a soft RICH identification of protons and kaons
\end{itemize}

\begin{figure}
 \mbox{
   \epsfxsize=8cm
   \epsffile{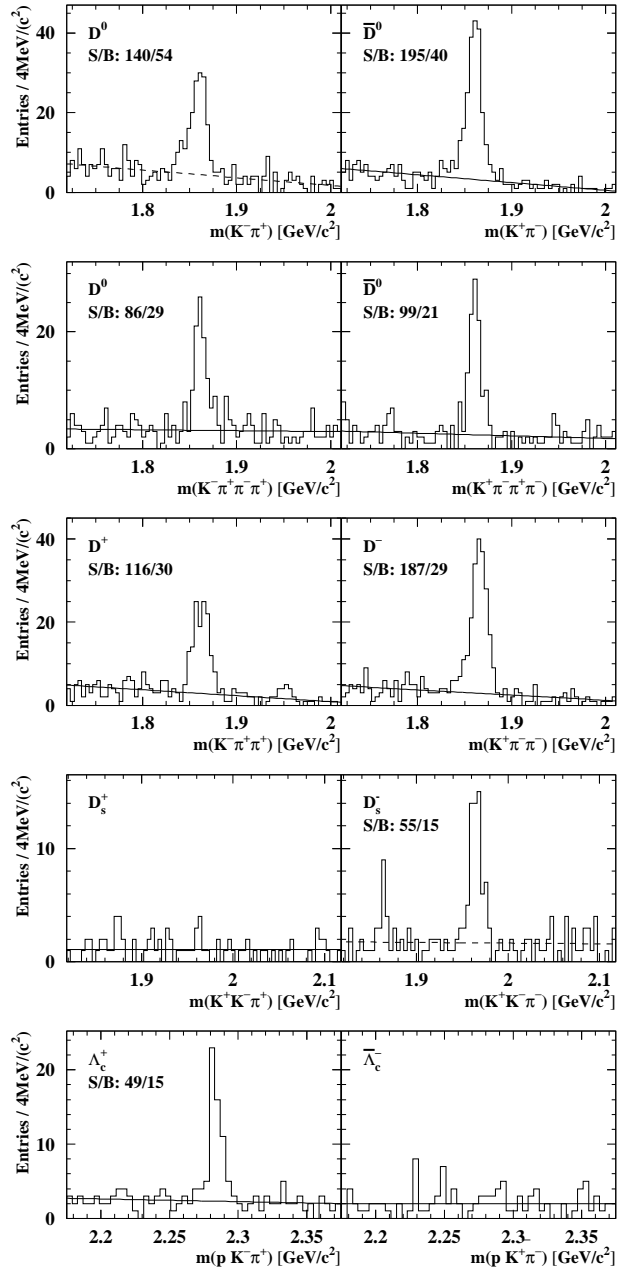}
   }
 \caption[]
 {Mass spectra in different decay channels 
  used for the cross section measurement.}
 \label{fig:signals}
\end{figure}

Table \ref{tab:channels} lists the reconstructed decay modes, the
number of reconstructed events and the respective branching ratios.
Figure \ref{fig:signals} shows the corresponding signals.

Acceptances and reconstruction efficiencies were determined using a detailed Monte Carlo 
simulation of the experiment. The charm events were simulated by an inclusive event 
generator which generated the charm (anti-charm) particle considered according to the 
measured kinematic distributions. The associated anti-charm (charm) particle was produced 
with the same kinematic distributions as the first particle, taking into account the 
azimuthal correlation and $\sum{p_t^2}$ distribution measured in a pion beam experiment 
\cite{beatrice:corr}. For the associated particles a mixture of 25\% of charged D-mesons, 
50 \% of neutral D-mesons and 25 \% of \PcgL was taken, in accordance with average yields 
of charmed particles measured in various experiments. The efficiencies were not sensitive 
to these percentages. The decay channel of the first particle was chosen explicitly while 
the decay of the associated charm particle was generated according to the known branching 
ratios. The remaining multiplicity of the event was generated by simulating the 
interaction of one hadron carrying the remaining momentum, using the {\sc Fritiof} 7.02 
\cite{fritiof} event generator for hadron-nucleus interactions. 
 The detector response was generated using {\sc Geant} 3.21. 
About 80000 events
were simulated per channel with different $x_F$ distributions to
ensure sufficient statistics at high $x_F$.

\begin{figure*}
\begin{center}
  \mbox{
     \epsfxsize=6.1cm
     \epsffile{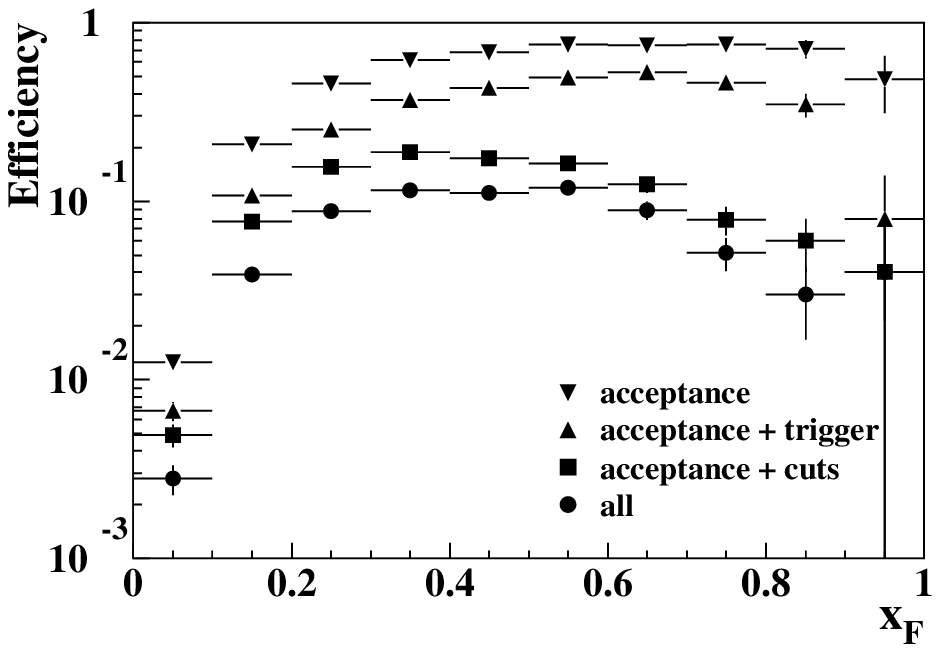}
     \hspace*{6mm}
     \epsfxsize=6.1cm
     \epsffile{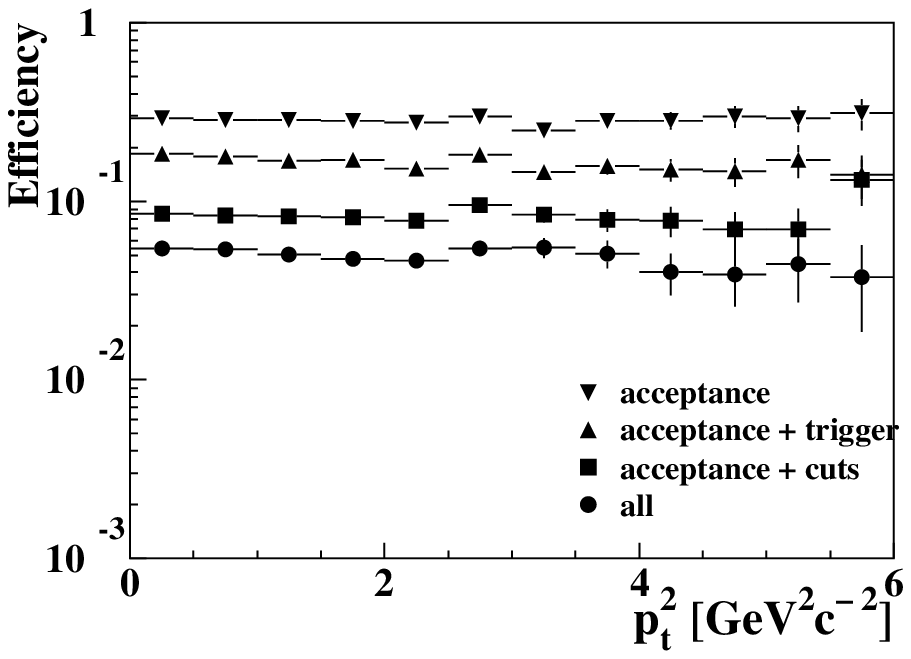} }
\end{center}
 \caption{ \PDm\ detection efficiency as a function of
   $x_F$ and $p_t$. ``cuts'' denotes the charm identification cuts.}
 \label{fig:eff}
\end{figure*}
\begin{figure*}
\begin{center}
  \mbox{
  \epsfxsize=6.1cm
     \epsffile{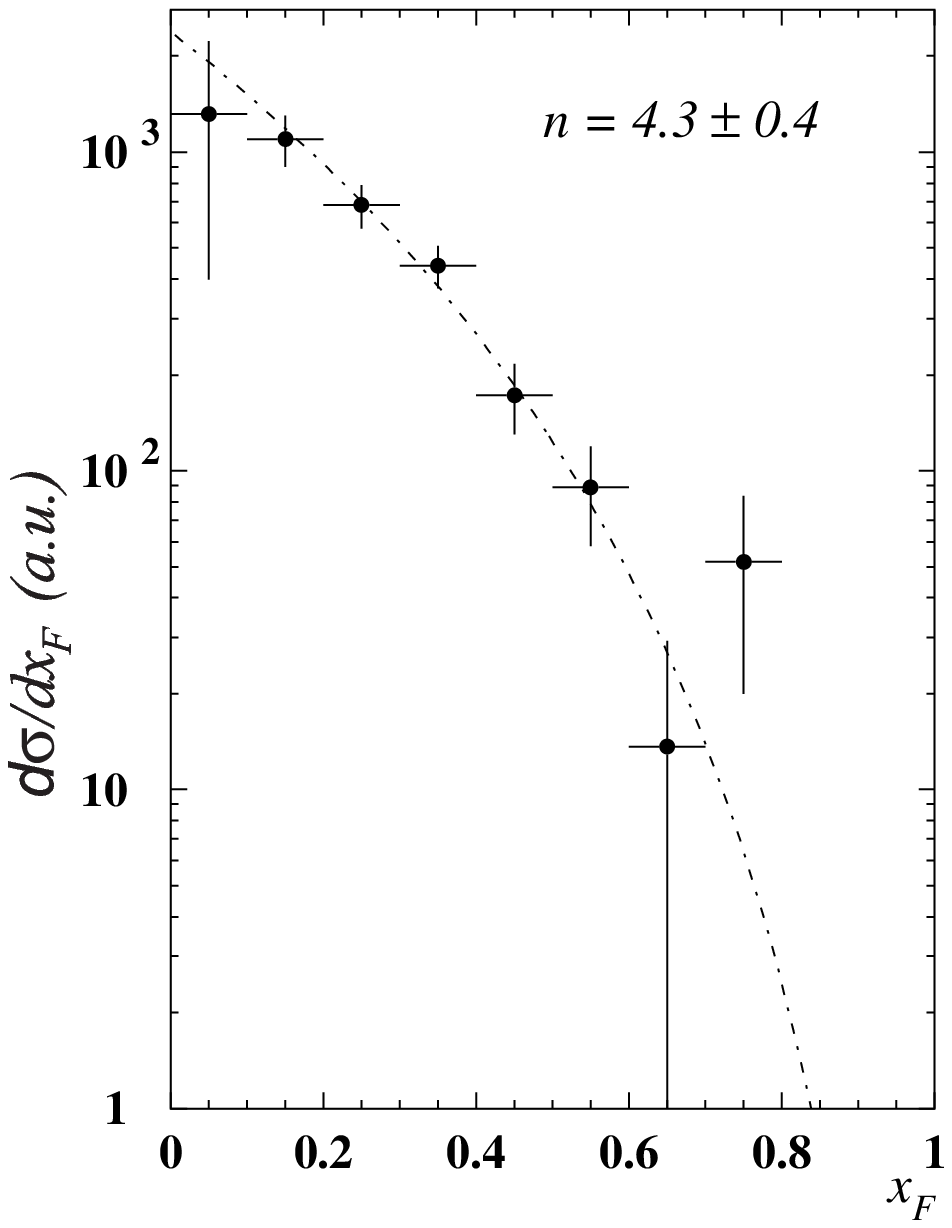}
     \hspace*{6mm}
     \epsfxsize=6.1cm
     \epsffile{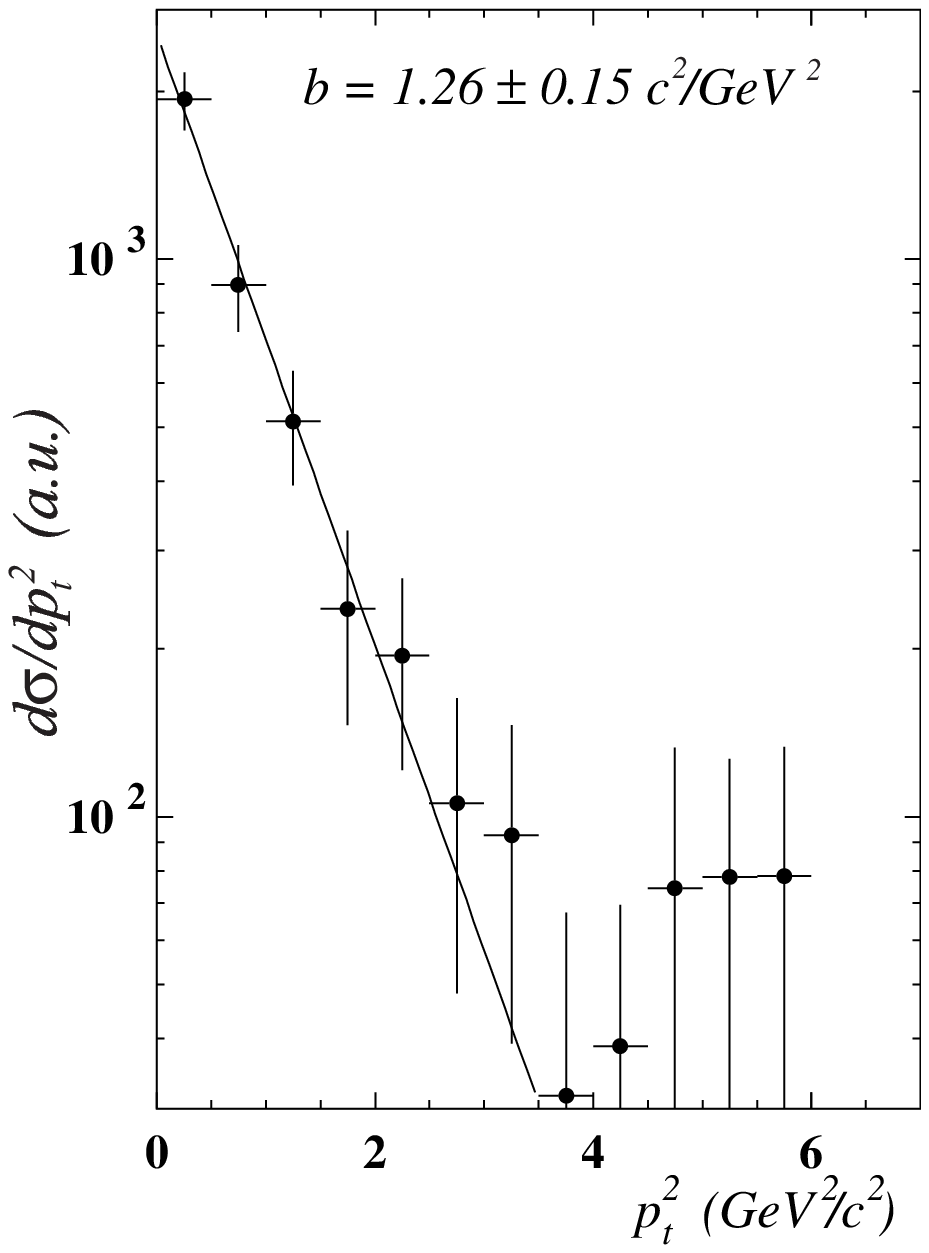} }
\end{center}
 \caption{ \PDm\ differential production cross sections $d\sigma /dx_F$
and $d\sigma /dp_t^2$ (arbitrary units). Fits as described in the text.}
  
 \label{fig:spectra}
\end{figure*}

The efficiencies obtained for the different decay channels analyzed show a similar 
behaviour. As an example, figure~\ref{fig:eff} shows the efficiencies for \PDm\ as 
functions of $p_t^2$ and $x_F$. While the efficiencies are  practically flat in $p_t^2$ 
over the whole relevant region, the dependence on $x_F$ shows a sharp rise at about $x_F 
= 0.1 $ and levels off above  $x_F = 0.2$ at a value of 10\% -- 20\%, depending on the 
decay channel. At high $x_F$, the charm identification cuts  produce a decrease in 
efficiency because of the difficulty to reconstruct a production vertex from the few 
remaining low momentum tracks, which often are not seen in the detector. 

After the  TRD decision in the trigger 
only $\approx2/3$ of the beam particles are
\PgSm . A large amount of contamination originates from pions from
\PgSm\ decays between the end of the beam collimator and the target. 
These were suppressed by a cut on the  track position/angle 
 correlation which is fulfilled
for beam particles coming from the production target, but not for decay pions (see 
\cite{beam} for details). Table \ref{tab:beam} shows the beam composition after this cut 
for both data taking periods. The remaining contributions are fast pions, misidentified 
as $\PgSm$ in the TRD, and a few \PKm\ and \PgXm\ which cannot be distinguished from 
\PgSm . 

\begin{table}[htb]
\begin{center}
  \begin{tabular}{|c|c|c|}
    \hline
    beam component & 1993     & 1994    \\    \hline
    \Pgpm & $0.14 \pm0.012$ & $0.21 \pm0.012$  \\
    \PKm  & $0.038\pm0.006$ & $0.048\pm0.008$  \\
    \PgXm & $0.013\pm0.001$ & $0.012\pm0.001$  \\ \hline
    \PgSm & $0.81\pm0.02$   & $0.73\pm0.02$    \\ \hline
  \end{tabular}
  \caption{Beam composition after elimination of \PgSm\ decays.}
  \label{tab:beam}
\end{center}
\end{table}

To correct for the beam contamination, we have to make assumptions about the  dependence 
of the charm production cross section on the beam projectile. In experiment E769 
\cite{Alves96b} the ratio of \PD/\PaD\ production by $\pi^-$ to production by protons was 
measured to be $r_{\pi p}=\sigma(\pi N\to \PD/\PaD ) / \sigma(pN\to \PD/\PaD )= 
1.31\pm0.24$ at 250 \gevc1 . With a 360 \gevc1 $\pi^-$ beam and a 400 \gevc1 proton beam, 
the LEBC-EHS (NA27) collaboration measured this ratio to be $r_{\pi p}=1.05\pm 0.20$ 
\cite{na27}. For the total charm production, a ratio $\sigma(\pi N\to c\overline{c} ) 
/\sigma(pN\to c\overline{c} ) = 1.00 \pm 0.15$ was obtained in  NLO and NNLO QCD 
calculations~\cite{frixione97,smith97}. We therefore correct for the pion and kaon 
contamination by using a value of  $r_{\pi p}=1.2\pm 0.3$, which amounts to a shift of 
$(4.5 \pm 7 )\% $. The possible effect from the $\Xi^-$ contamination is negligible in 
this context.

\section{Cross Section Results} 

Since the detector acceptance drops sharply at small values of $x_F$, the measured 
differential cross sections were extrapolated to $x_F = 0$ using the usual ansatz 
$d\sigma/dx_F \propto (1-x_F)^n$ where $n$ was obtained from a fit to the data above $x_F 
= 0.1$. As an example, the measured   $d\sigma/dx_F$ distribution for \PDm\ production is 
shown in figure~\ref{fig:spectra} together with the fit results. The values of n obtained 
in the fits are included in table \ref{tab:channels}. The errors quoted are the 
statistical errors of the fits only. The values of n obtained are in agreement with 
values obtained in pion and proton beam experiments (\cite{beatrice,Alves96b} and earlier 
references therein). The errors preclude any conclusions about possible differences in n 
between the different charm particles produced in this experiment or between our \PgSm\ 
beam result and the pion or proton beam results. 

We also fitted the observed p$_t$ spectra with a form $d\sigma /dp_t^2  \propto 
exp(-b\cdot p_t^2)$, assuming b to be independent of n. Figure~\ref{fig:spectra} shows 
the spectrum and fit result for \PDm\ production, and the values of b obtained are listed 
in the last column of table \ref{tab:channels}. Again we observe agreement with the 
earlier experiments quoted above, but are unable to draw further conclusions. 
 
The data from both targets were  added and extrapolated from \PgSm\ -nucleus to \PgSm\ 
-nucleon cross sections assuming an $A^{\alpha}$ dependence with $\alpha$=1. The same 
dependence was used in the analyses of the proton beam experiments listed in 
table~\ref{tab:allcross}. The most precise measurements of the exponent $\alpha$ in charm 
production come from the pion beam experiments 
 WA82 \cite{wa82alpha}, E769 \cite{e769alpha} and
BEATRICE(WA92) \cite{beatrice}. The averages from the three experiments are 
$\alpha=0.95\pm 0.04$ for both \PDz/\PaDz\ and \PDpm\ production, with no significant 
dependence of $\alpha$ on the longitudinal or transverse momenta. Use of $\alpha=0.95$ 
would increase the resulting \PgSm\ -nucleon cross sections by a factor of 1.13 (1.23) 
for carbon (copper) targets. 
 
\begin{table} [htb]
\begin{center}
\begin{tabular}[h]{|rcl|c|c|}
 \hline
 \multicolumn{3}{|c|}{ Decay} &  $\sigma(\cbar) [\mu\mbox{b}]$ &
                               $\sigma(\cq)   [\mu\mbox{b}]$  \\ &&&&\\
 \hline
 \PDz   & \dcy &\PK \Pgp  &  2.93 $\pm$ 0.29 &  2.35 $\pm$ 0.33  \\
        & \dcy &\PK \Pgp \Pgp \Pgp &  2.07 $\pm$ 0.41 &  2.38 $\pm$ 0.52  \\
        & \multicolumn{2}{c|}{mean} & 2.64 $\pm$ 0.24 &  2.36 $\pm$ 0.28  \\
 \PDpm  & \dcy &\PK \Pgp \Pgp &  1.30 $\pm$ 0.14 &   1.13 $\pm$  0.16 \\
 \PsDpm & \dcy &\PK \PK \Pgp  &  1.60 $\pm$ 0.42  &               \\
 \PcgL  & \dcy &\PK \Pp \Pgp  &                   &  2.37 $\pm$  0.68 \\ \hline
\multicolumn{3}{|c|}{ Sum}    &  5.54  $\pm$ 0.51   &  5.86  $\pm$ 0.75 \\ \hline
\end{tabular}
\caption
  {Production cross sections for anti-charmed and charmed hadrons
   in the range $x_F>0$ .
  The third line gives the \PDz\ cross sections 
  averaged from the measurements in the \PK \Pgp\ and \PK \Pgp \Pgp \Pgp\
  decay channels.   
   Errors are statistical only. } 
\label{tab:ourcross}
\end{center}
\end{table}

The resulting integrated cross sections 
are listed in table \ref{tab:ourcross}.
A first consistency check of the analysis can be obtained by
comparing the results from the two-body and four-body 
decay modes of the \PDz, which
are in good agreement.

The total $c\bar{c}$ production cross section is taken to be the mean of the sums of the 
observed charm and anticharm cross sections (using a weighted mean for the two \PDz\ 
cross sections) with the result: $\sigma_{\cq\cbar}(x_F > 0) = 5.6\, \pm\, 0.4\,\, 
(stat)\, \mu b $. 

Extensive studies of possible systematic effects have been carried out. No considerable 
dependence of the result on the cut values or on reasonable variations of detector 
efficiencies was found. We assign an overall systematic error of $20\%$ to all cross 
sections due to the uncertainties of the Monte Carlo simulation, the extrapolation to 
$x_F$=0 and the luminosity measurement. 

Our cross section result  is based on the observation of \PD\ and ${\rm D_s}$ mesons, to 
which all excited charm mesons will eventually decay, and of \PcgL , to which all charmed 
non-strange baryons will decay. Therefore possible undetermined contributions from charm 
anti-baryons and from charmed-strange baryons remain, which will be discussed in the 
following. 

A significant contribution to the total cross section by charm anti-baryons is not 
expected due to the observed strong asymmetry in \PcgL/\PacgL\ production \cite{asym}. 
From the results of a direct search for charmed-strange baryon production, we cannot 
exclude production cross sections of  $1\,\mu\mbox{b}$ \cite{theses}. We therefore add a 
further systematic uncertainty of $+1\,\mu\mbox{b}$ from \PcgX\ production to our total 
$c\bar{c}$ production cross section. A contribution from  \PcgO\ should be small compared 
to the overall uncertainties. 

\noindent The total $c\bar{c}$ production cross section for $x_F>0$ then is:\\ 
\centerline{ $\sigma_{\cq\cbar}(x_F>0) = 5.6\, \pm\, 0.4\, (stat)\, \pm\, 1.1\, (syst)\, 
+1.0\, (\PcgX )\,\, \mu\mbox{b} $}

\begin{table*}[t]
\begin{center}
\begin{tabular}[h]{|l|r|l|cc|c|l|}
 \hline
 Experiment & beam, & targets &$\sigma (\PD /\PaD )$ & x$_F$ range
 & $\sigma ( \cq \cbar )$ for x$_F>0$ \\
  & mom. & &  $[\mu\mbox{b} ]$ & & $[\mu\mbox{b} ]$ \\ \hline
 NA32 \cite{accmor} & p, 200 & Si &
    1.5$\pm$0.7 & $>$0 & 2.25$\pm$1.10 \\ \hline
 E769 \cite{Alves96b} & p, 250 & Be,Al,Cu,W &
   9.0$\pm$1.5 & $>$0 & 13.5$\pm$2.9 \\ \hline
 NA27 \cite{na27} & p, 400 & p & 30.2$\pm$2.9 & all & 22.6$\pm$3.8 \\  \hline
 E653 \cite{e653} & p, 800 & emulsion & 
     76$\pm$20 & all & 57$\pm$17 \\ \hline
 E743 \cite{e743} & p, 800 & p & 48$\pm$12 & all & 36$\pm$10 \\ \hline
 WA89 & $\PgSm$ , 340 & C,Cu &
     7.4$\pm$1.5 &  $>$0 & 11.3$\pm$2.4+2.0 \\ \hline
\end{tabular}
\caption
  {Production cross sections measured in proton beams and in this expt.
   Beam momenta are given in \gevc1 .}
\label{tab:allcross}
\end{center}
\end{table*}

\begin{figure}
\begin{center}
 \mbox{
     \epsfxsize=8cm
     \epsffile{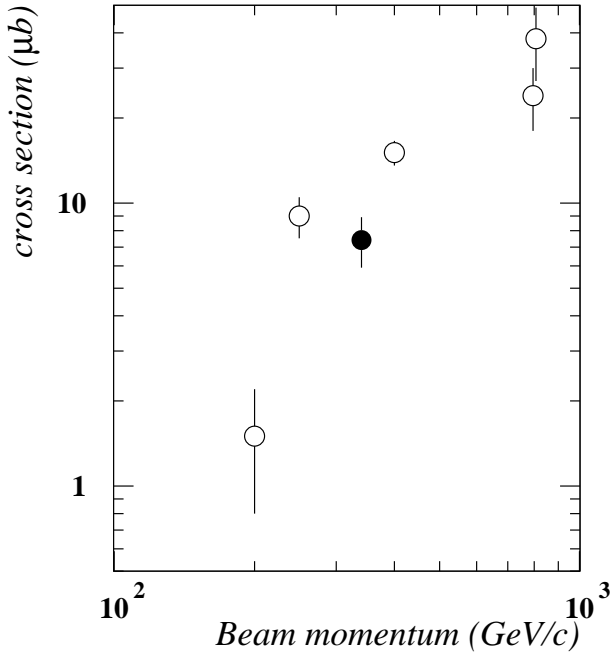} }
 \end{center}
 \caption{\PD /\PaD\ production cross sections 
   by $\PgSm$ (this expt., full circle) and protons (open circles).}
 \label{fig:dcross}
\end{figure}

\section{Discussion}

In the hyperon beam experiment WA89 at CERN, close to 1000 charm particles were found in 
340 \gevc1 $\PgSm$-nucleus interactions. We determined production cross sections for \PDz 
, \PaDz , \PDpm , \PsDm\ and \PcgLp\ in the region $x_F>0$. While a clean \PcgLp\ signal 
could be observed, the \PacgL\ could not be singled out. This suppression of antibaryon 
production in baryon beams is well known from non-charm baryon production. The cross 
sections for \PDpm\ and \PD /\PaD\ production are comparable. The suppression of 
$\mbox{D}_s^+$ production relative to  $\mbox{D}_s^-$ is probably caused by the 
suppression of $\overline{s}$ production relative to $\overline{u},\overline{d}$ 
production, while the $s$ quark of the $\mbox{D}_s^-$ can be supplied by the \PgSm\ beam 
particle. From this argument one would expect an enhancement of \PDm\ over \PDp\ 
production as well, but this enhancement is not so strong since light antiquarks are 
produced more copiously than the heavier $\overline{s}$ quarks. Indeed, we observed a 
significant $\rm{ D^- /D^+}$ production asymmetry at larger $x_F$ \cite{asym}, but the 
effect on the integrated cross section for $x_F >0$ is not observable within the errors 
of our measurements.

In figure ~\ref{fig:dcross} and in table~\ref{tab:allcross} our result
for the (\PD\ or \PaD ) production cross section 
$\sigma_{\PD /\PaD}(x_F >0 )= 7.4 \pm  0.4(stat.)\pm 1.5(syst.)\, \mu b$
is compared to measurements in proton beams
\cite{Alves96b,na27,accmor,e653,e743}.
Proton beam measurements valid for $-1<x_F<1$ have been divided by
2 for this comparison. Our result is significantly lower than the
proton beam values. It is not clear, however, whether for this comparison
we should include  all or part of our measured value for \PsDm\ production,
$\sigma_{D_s^-}\, = 1.6\, \pm 0.4(stat.) \pm
0.3(syst.)\, \mu b$, since our beam projectile carries a strange quark.

Our measured value for the ratio of charged and neutral \PD/\PaD\
production,
$\sigma(\PDpm )/\sigma(\PDz,\PaDz )=0.49\pm0.06$,
is compatible with the mean value from the
proton beam experiments  
\cite{Alves96b,na27,e653,e743}, 
$\sigma(\PDpm )/\sigma(\PDz,\PaDz )=0.66\pm0.09$.
The corresponding ratio from pion beam experiments at beam
energies between 200 and 600 \gevc1 is
$\sigma(\PDpm ) /$ $ \sigma(\PDz,\PaDz )=0.415\pm0.010$
in good agreement with our result
(for the pion beam results,
see ref.~\cite{beatrice} and earlier references therein).
For the ratio of \PsDm\  and
\PD\ production we obtained
$\sigma(\PsDm ) / \sigma(\PD,\PaD )=0.22\pm0.06$.
This ratio has not been measured in proton beams.
In the pion beam experiments quoted above, 
the ratio of \PsDp\ and \PsDm\ production to \PD\ 
production has been measured to be 
$\sigma(\PsDpm )/\sigma(\PD,\PaD )=0.129\pm0.012$.
Since both \PsDp\ and \PsDm\ are non-leading when produced in
a pion beam, we can safely assume that 
$\sigma(\PsDm )/\sigma(\PD,\PaD )=0.065\pm0.006$
in pion beams, which is markedly lower than our \PgSm\ beam result.
This is another clear manifestation of the leading particle effect.

\begin{figure}
\begin{center}
 \mbox{ 
     \epsfxsize=8cm
     \epsffile{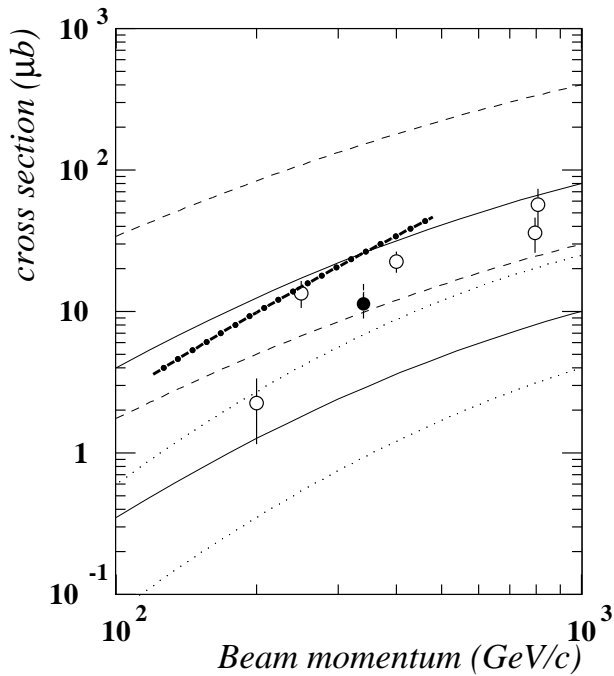} }
 \end{center}
 \caption{Total charm production cross sections 
   by $\PgSm$ (this expt., full circle) and protons (open circles)
   compared to theoretical predictions    as explained in the text.}
 \label{fig:sigtot}
\end{figure}

Theoretical calculations exist only for the total charm
production cross section, $\sigma_{\cq\cbar}$, for all x$_F$.
Therefore our result and the 
results of the proton beam experiments~\cite{Alves96b,accmor},
obtained in the kinematic region x$_F>0$,
were multiplied by a factor of 2.
This factor of 2 holds for interactions
with equal beam and target particles, but even in
 \PgSm\ N interactions,  no significant deviation 
from 2 is expected, since the production cross sections integrated over the range 
$0<x_F<1$ show only small differences between leading and non-leading charm hadrons as 
discussed above. Furthermore, the \PD/\PaD\ production cross sections measured in the 
proton beam experiments have to be extrapolated to  total charm production. Our value for 
the ratio of total charm production to  \PD /\PaD\ production is $\sigma_{\cq 
/\cbar}/\sigma_{\PD /\PaD}=1.53\pm 0.15 + 0.10$. The  E769 collaboration \cite{Alves96b} 
measured $\sigma_{\PD +\PaD}=9.0\pm 1.5 \mu b$ and $5.1<\sigma_{\PcgL}<21.5 \mu b$ in 
their proton beam, which is consistent with our ratio. In their pion beam they measured  
$\sigma(\PcgL + \rm{D}_s  ) /\sigma(\PDp + \PDz) = 0.53 \pm 0.14$ (particles and 
antiparticles added). In the photoproduction experiment NA14/2 this ratio was measured to 
be 0.36$\pm$0.11 \cite{na14/2}. 
 These values indicate that the ratio 
$\sigma_{\cq/\cbar}/\sigma_{\PD /\PaD}$ does not depend strongly on the nature of the 
beam particle. We used  a value $\sigma_{\cq /\cbar}/\sigma_{\PD /\PaD}= 1.5\pm 0.2$ to 
calculate $\sigma_{\cq /\cbar}$ from the proton beam results on $\sigma_{\PD /\PaD}$. 
Finally, the production cross section for {\it $c\overline{c}$-pairs}, 
 $\sigma_{\cq\cbar}$, was obtained by dividing the
above results for {\it $c$ or $\overline{c}$} by a factor of 2. The results are listed in 
the last column of table~\ref{tab:allcross} and are compared in figure~\ref{fig:sigtot} 
with theoretical calculations. The dashed/solid/dotted bands were obtained by Frixione et 
al. in NLO QCD-calculations for charm quark mass values of 1.2/1.5/1.8 \Gevc2 , 
respectively \cite{frixione97}. The width of the bands for fixed  $m_c$ is given mainly 
by the uncertainties of the renormalization scales. The proton beam measurements prefer 
the upper range of the predictions, and seem to exclude  a value of $m_c=1.8 \Gevc2$. 
This holds also for charm production measurements in pion beams, as discussed in 
\cite{frixione97,frixione94}. The thick dot-dashed line denotes the results of Smith and 
Vogt \cite{smith97} for a charm quark mass of 1.5 \Gevc2 . These authors used resummation 
techniques and approximations for NNLO. At our beam energy, their results are larger by a 
factor 3 than the NLO calculations for the same charm quark mass.

Our experiment provides the first comparison of charm production by \PgSm\ and by 
protons. Our result seems to be lower than the proton beam results, in line with the 
observation that the total \PgSm\ N cross sections are smaller than the pp cross sections 
\cite{sigma-cross}. However, in view of the problems encountered when comparing cross 
sections from different experiments, it may be premature to claim that 
$\sigma_{\cq\cbar}$ or $\sigma_{\PD /\PaD}$ is smaller in \PgSm\ than in proton beams. 
The forthcoming results from the hyperon beam experiment SELEX at FNAL will help to 
clarify this point \cite{selex}.


\end{document}